\documentclass{aa}
\usepackage{graphicx}
\usepackage[]{natbib}
\def\lunits{$\rm erg\,s^{-1}$~}
\def\funits{$\rm erg\,cm^{-2}\,s^{-1}$~}
\def\cunits{$\rm cm^{-2}~$}

\def\chandra{{\it Chandra~}}

\begin{document}
 \title{The Compton-thick AGN in the CDFN}

%   \subtitle{ }

  \titlerunning{Compton-thick AGN in the CDFN}
    \authorrunning{I. Georgantopoulos et al.}

   \author{I. Georgantopoulos\inst{1},
           A. Akylas \inst{1},
           A. Georgakakis \inst{1},
           M. Rowan-Robinson \inst{2} 
           }

   \offprints{I. Georgantopoulos, \email{ig@astro.noa.gr}}

   \institute{Institute of Astronomy \& Astrophysics,
              National Observatory of Athens, 
 	      Palaia Penteli, 15236, Athens, Greece \\
              \and
         Astrophysics Group, Blackett Laboratory, Imperial College,  
       Prince Consort Road, SW7 2BZ, U.K. \\
             }

   \date{Received ; accepted }

\abstract{We present X-ray spectral analysis of the brightest sources 
 ($\rm f_{2-10~ keV}>10^{-15}$ \funits) in the Chandra Deep Field North.
 Our sample consists of 222 sources; for the vast majority     
 (171) either a spectroscopic or a photometric redshift is available.  Our goal is to discover the 
 Compton-thick AGN in a  direct way i.e. through their X-ray spectra. Compton-thick AGN give away their presence 
  in X-rays either directly through the absorption turnover redshifted in the Chandra passband, 
   or through a flat, reflection-dominated, spectrum.  The above selection criteria yield 10 Compton-thick AGN candidates of which the nine 
    are reflection dominated.  
    The IR or sub-mm data where available, corroborate the presence of a heavily obscured 
     nucleus in most cases. 
       All the five candidate Compton-thick sources with available 24$\mu m$ data present very high values 
    of the $\rm f_{24}/f_R$ flux ratio suggesting that they are dust obscured galaxies. 
      The low $\rm f_x/f_{IR}$ ratio also suggest the presence of obscured nuclei in many cases.
   Four of the candidate Compton-thick sources  are associated with sub-mm galaxies  at high redshifts z$\sim2$. 
      The number count vs. flux distribution of the candidate Compton-thick AGN as well 
     as their distribution with redshift  agree  reasonably well with the predictions of the 
      X-ray background synthesis models of Gilli et al.  \keywords {X-rays: general; X-rays: diffuse emission;
X-rays: galaxies; Infrared: galaxies}}
   \maketitle
%
%________________________________________________________________

\section{Introduction} 

The hard X-rays (2-10 keV) present the advantage that they can penetrate large amounts of
  interstellar gas and thus can detect AGN which would be missed in optical wavelengths. The High Redshift 
Universe has been probed at unparallel depth with the deepest ever observations  in 
the Chandra Deep Field North and South (Alexander et al. 2003, Giacconi et al. 2002, 
 Luo et al. 2008). These observations resolved 80-90\% of the extragalactic 
X-ray light, the X-ray background, in the 2-10 keV band revealing a sky density of about 5000 
sources per square degree (Bauer et al. 2004), the vast majority of which are AGN (for a review see Brandt 
\&  Hasinger 2005). 
 In contrast, the optical surveys for 
QSOs (eg the COMBO-17 survey or the 2QZ) reach a surface density of about an order of magnitude lower 
(Wolf et al. 2003, Croom et al. 2003). This immediately demonstrates the power of X-ray surveys for detecting AGN 
and thus for providing us with the most unbiased view of the accretion history of the Universe.  
 
However, even the extremely efficient X-ray surveys may be missing a fraction of heavily obscured 
sources. This is because at very high obscuring column densities ($10^{24}$ \cunits) 
around the nucleus even the hard X-rays (2-10 keV) are significantly suppressed. These are the so called Compton- 
thick AGN (see Comastri 2004 for a review) where the probability for Thomson scattering becomes significant.
  The X-ray background 
synthesis models (Comastri et al. 1995, Gilli et al. 2007) can explain the peak of the X-ray background at 40 keV, where most of its energy 
density lies, (eg Frontera et al. 2007, Churazov et al. 2007) only by invoking a numerous 
population of Compton-thick sources. 
However, the exact surface density of Compton-thick AGN required is still debatable 
 (see e.g. Sazonov et al. 2008, Treister et al. 2009). 
Additional evidence for the presence of an appreciable Compton-thick 
population comes from the directly measured space density of black holes in the local Universe. It 
is found that this space density is a factor of two higher than that predicted from the X-ray 
luminosity function (Marconi et al. 2004). This immediately suggests that the X-ray luminosity 
function is missing a large number of AGN.    
  
  In recent years there have been many efforts to discover Compton-thick AGN 
   in the local Universe  by examining optically selected, AGN samples classified on the basis of 
   narrow  emission line diagnostic ratios (Risaliti et al. 1999, Cappi et al. 2005,
   Akylas \& Georgantopoulos 2009).  This is motivated by the fact that the narrow emission line region, 
  which represents an isotropic AGN property, is an excellent proxy of the power of the nucleus.  
   The advent of the {\it SWIFT}  (Gehrels et al. 2004) and the {\it INTEGRAL} missions (Winkler et al. 2003) which carry ultra-hard X-ray 
    detectors ($>$15 keV), with  limited imaging capabilities, helped 
     towards further constraining the number density of Compton-thick sources 
      at very bright fluxes, $\rm f_{17-60 keV} >10^{-11}$ \funits in the local Universe, $z<0.1$
        (Beckmann et al. 2006, Bassani et al. 2006, Sazonov et al. 2007, Winter et al. 2008, Winter et al. 2009, Tueller et al. 2009).    
   
 Mid-IR surveys (8-60 $\mu m$) have been used as an alternative tool to 
  identify Compton-thick sources.  
 This is because the absorbed optical and UV radiation heats the dust and is re-emitted 
  at IR wavelengths. Therefore, {\it Spitzer} surveys provide a promising technique to 
   detect Compton-thick sources.  The difficulty faced in mid-IR surveys is that the
AGN are vastly outnumbered by normal galaxies 
   and some selection method is necessary to separate the two populations.
   Selection criteria which use Spitzer/IRAC mid-IR colours
(Lacy et al. 2004; Stern et al. 2005) appear to be more prone to residual 
    galaxy contamination at faint optical magnitudes or faint X-ray fluxes.  
    Mid-IR Spectral Energy Distribution 
    (SED)  techniques (Alonso-Herrero et al. 2006;  Poletta et al. 2006; Donley et al. 2007)
     are much more successful in picking out  AGN but not necessarily the bulk 
      of the Compton-thick population. Georgantopoulos 
      et al. (2008) compare in detail the above methods using
       X-ray and mid-IR  data in the Chandra Deep Field North and discuss their efficiency for finding  Compton-thick sources.
       Alternatively, methods which use a combination of optical and mid-IR photometry
        have been proposed (eg Daddi et al. 2007, Fiore et al. 2008). 
         The method of Daddi et al. (2007) involves the selection of mid-IR excess sources.     
         Fiore et al. (2008)  instead select the optically faint, mid-IR selected sources  
          (see Houck et al. 2005) which have additionally red optical colours.  These techniques  may provide a 
         much more efficient tool for unearthing heavily obscured sources
      {\it below the flux limit} of the current Chandra surveys. Indeed, the stacked X-ray 
       signal of these sources appears to be flat indicative 
        of absorbed sources,    
         Fiore et al. (2008), Georgantopoulos et al. (2008), Fiore et al. (2009),
         in the case of the CDF-S, CDF-S and COSMOS fields respectively;   
         but see also  Pope et al. (2008) who argue that the galaxy contamination 
           may still be significant.  
            In any case, it is impossible to argue unambiguously that these sources 
             are Compton-thick given that only a stacked hardness ratio is available
              and not individual good quality X-ray spectra.             
   
 Here instead, we focus on identifying  the Compton-thick sources 
 which are   present among the  {\it detected  sources} in deep X-ray surveys.
As Compton-thick  sources have a quite distinctive X-ray spectrum, i.e. either a spectral turnover 
 in the transmission dominated case or a flat continuum in the reflection-dominated case, 
  X-ray spectroscopy provides  a reasonably robust way for identifying these heavily obscured sources.
  In contrast, methods which are based only on  IR  diagnostics i.e. either spectroscopy or Spectral 
   Energy Distribution  fitting (e.g. Alexander et al. 2008) may be more prone to  uncertainties. 
    For example, even  the nearby Compton-thick AGN 
    NGC6240 is classified as a star-forming galaxy or a LINER on the basis of mid-IR diagnostics (e.g. Lutz et al. 1999).    
 The X-ray spectroscopy technique has been previously applied in the 1Ms Chandra Deep field South  by 
  Tozzi et al. (2006) and Georgantopoulos et al. (2007). Several Compton-thick candidates have been 
   identified. However, the limited photon statistics hampered the identification 
    of bona-fide reflection-dominated (flat-spectrum) AGN.  
  To remedy this, we explore only the X-ray spectral properties of the bright sources ($\rm f_{2-10 ~keV}>10^{-15}$ \funits) 
  in the Chandra Deep Field North. Our goal is to find the sources which either show 
  a rest-frame column density of $\rm N_H>10^{24}$ \cunits 
   or alternatively a flat spectrum $\Gamma <1.4 $ at the 90\% confidence level).      
    We adopt $\rm H_o
=  75  \,  km  \,  s^{-1}  \,  Mpc^{-1}$,  $\rm \Omega_{M}  =  0.3$,\
$\Omega_\Lambda = 0.7$ throughout the paper.

\section{Data}
 
 \subsection{CDFN}
The CDF-N is centred at $\alpha = 12^h 36^m
49^s.4$,  $\delta =  +62^\circ 12^{\prime}  58^{\prime\prime}$ (J2000)
and has been surveyed extensively  over a range of wavelengths by both
ground-based facilities and space  missions. The multiwaveband data in
this  field include  {\it Chandra}  X-ray observations,  {\it Spitzer}
mid-IR photometry, {\it HST}/ACS high resolution optical imaging, deep
optical  photometry and  spectroscopy  from the  largest ground  based
telescopes.

The 2Ms  {\it Chandra} survey of  the CDF-N consists  of 20 individual
ACIS-I (Advanced CCD  Imaging Spectrometer) pointings observed between
1999 and 2002.  The combined observations cover a total area of $447.8
\,  \rm  arcmin^2$ and  provide  the  deepest  X-ray sample  currently
available together with the Chandra Deep Field South (Luo et al. 2008).   
  Here,   we   use    the   X-ray   source   catalogue   of
 (Alexander et al. 2003), which  consists of  503 sources detected  in at
least one of  the seven X-ray spectral bands  defined by these authors
in the  range $0.3-10$\,keV.  The flux  limit in the
$2-10$\,keV   band is $1.4\times
10^{-16}$\,\funits.  The Galactic column density towards
the CDF-N is $1.6\times 10^{20}$\,\cunits (Dickey \& Lockman 1990).

The central region of the CDF-N has been
observed in the mid-IR by the {\it Spitzer} mission  (Werner et al. 2004)
as part of  the Great Observatory Origin Deep Survey (GOODS).  These observations cover an  area of about $10
\times 16.5\rm  \, arcmin^2$  in the CDF-N  using both the  IRAC (3.6,
4.5,  5.8 and  8.0$\, \rm  \mu m$)  and the  MIPS ($24\,  \rm  \mu m$)
instruments onboard Spitzer.   Here we use the 2nd  data release (DR2)
of  the IRAC  super-deep images  (version 0.30)  and the  interim data
release  (DR1+) of the  MIPS 24$\,  \rm \mu  m$ mosaic  (version 0.36)
provided by the  GOODS team (Dickinson et al.   2003). 
 Sources  are  detected in  these  images using  the
SExtractor  (Bertin \& Arnouts 1996) software.  Full  details on  the source
extraction     and     flux     derivation    are     presented     in
(Georgakakis et al. 2007).  The
$\rm 24 \mu m$ selected sample consists of 1619 sources to the
flux density limit of about 15\,$\mu$Jy.

Multi-waveband optical imaging ({\it  UBVRIz}) in the CDF-N region has
been  obtained  using the SUBARU 8.2-m telescope by  (Capak et al. 2004).   
 Here,  we  use  the  $R$-band
selected  sample, which  contains  47451 sources  down  to a  limiting
magnitude of  $R_{\rm AB}=26.6$\,mag (5$\sigma$).   These observations
cover about  $\rm 0.2\,  deg^2$ and extend  beyond the GOODS  field of
view  ($\approx \rm  0.05\,deg^2$). The  1619 $\rm 24  \mu m$  sources are
first cross-correlated with the IRAC $3.6\mu m$ catalogue using a matching radius 
 of 2 arcsec. Then  we cross-correlate with the  
(Capak et al. 2004) $R$-band catalogue using a
matching radius  of 1.5 arcsec. Optical  identifications are available
for 1409 out of the 1619 sources (87\%). 
Finally, we cross-correlate between the X-ray position and the R-band 
 positions using a 2 arcsec radius.

\subsection{The X-ray sample}
\begin{figure}
   \begin{center}
\includegraphics[width=8.5cm]{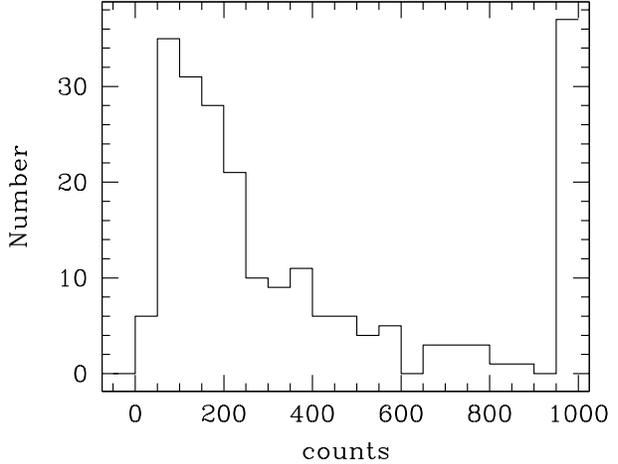} 
\caption{ The photon count distribution for the 222 sources with 
 $\rm f_{2-10~ keV} > 10^{-15}$ \funits. The last bin includes all sources 
  with over 1000 photons.}
  \label{counts}
  \end{center}
\end{figure}

We confine our analysis  to the bright sources  ($\rm f(2-10 ~ keV) > 10^{-15}$ \funits) in the 
 2-10 keV band. 
 The choice of this particular flux limit  is dictated by  a balance between the 
  number of Compton-thick sources expected (Gilli et al. 2007) and  the need 
   for good photon statistics. 
    There are 222 sources brighter than this flux limit. 
    Their photon count distribution is given in Fig. \ref{counts}. 
    
     107 sources have a spectroscopic redshift 
     available (Barger et al. 2003). For another 64 sources, photometric redshifts
      have been derived in Georgantopoulos et al. (2008)  
      using the code of Babbedge et al. (2004).
      The full description of the photometric redshift analysis is given in Georgantopoulos et al. (2008).
       Here, we  briefly mention that the photometric redhshifts are obtained by fitting the U-band 
        to 4.5$\mu m$ photometric data with a library of 8 templates which include  both galaxies and AGN.
         The $1\sigma$ photometric redshift accuracy is $\delta z/ (1+ z_{spec}) \approx 0.04 $.  
          At a second step, the mid-IR SED is fit, after subtracting the stellar contribution by extrapolating the best-fit 
          galaxy template determined in the previous step. The mid-IR SED modeling provides an estimate of the total infrared luminosity in the wavelength range 3-1000 $\mu m$ (see  Rowan-Robinson et al. 2005). This luminosity can be considered a  good proxy of the bolometric luminosity within a factor of  two.  
 Finally, 51 sources have no secure optical counterpart and thus no photometric redshift available.

\section{Analysis}

\subsection{X-ray Spectral Fits} 

\begin{figure}
\includegraphics[width=8.5cm]{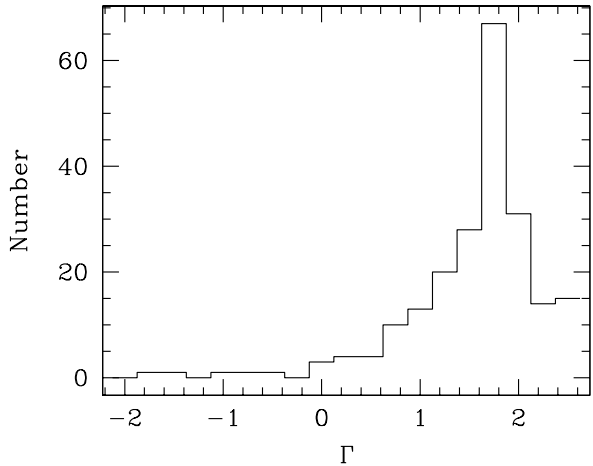} 
\caption{ The distribution of photon indices $\Gamma$ in our spectral fits}
  \label{gamma}
\end{figure}

We use the {\sl SPECEXTRACT} script in the CIAO v4.2 software 
 package to extract spectra from the 20 individual CDF-N observations.
  The extraction radius varies between 2 and 4 arcsec with increasing off-axis angle. 
   At low off-axis angles ($<$4 arcsec)  this encircles 90\% of the light at an energy of 1.5 keV.  
 The same script extracts response and auxilliary files. 
  The addition of the spectral, response and  auxiliary files has been 
   performed with the FTOOL  tasks {\sl MATHPHA}, {\sl ADDRMF} and {\sl ADDARF} respectively.  
 The data are grouped using the CIAO {\sl DMGROUP} task so that there are 15 counts per bin and thus 
  $\chi^2$ statistics can apply. 
   For a small number of sources (six)
which have limited photon statistics, we use
the C-statistic technique (Cash 1979) specifically developed
to extract spectral information from data with low
signal-to-noise ratio. 
 We use the XSPEC v12.4 software
package for the spectral fits (Arnaud 1996).  We fit the data using a power-law
model absorbed by two cold absorbers: wa*zwa*po in
XSPEC notation. The first column is fixed to the Galactic (Dickey \& Lockman 1990)
 while the second one is the rest-frame
intrinsic column density. We treat both the intrinsic column density 
and the photon index  as free parameters. 
The distribution of the photon indices is given in Fig. \ref{gamma}. 
       
%%%%%%%%%%       CT  TABLE %%%%%%%%%%%%%%%%%%%%%%%%%%%%%%
\begin{table*}
\begin{center}
\caption{X-ray spectra of the candidate Compton-thick AGN}
\label{tab_ct}
\begin{tabular}{lccccccccc}
\hline 
\multicolumn{3}{l}{} & \multicolumn{3}{c}{Power-law} & \multicolumn{2}{c}{Reflection} & \multicolumn{2}{c}{}  \\
ID  & $\alpha, \delta$ & z &  $N_H$      & $\Gamma$    & $\chi^2/  \nu $   & $\Gamma$  & $\chi^2/ \nu$  & log$L_x$  & flux \\ 
(1)   & (2)&    (3)        &  (4)          &  (5)         &   (6)       & (7)        & (8) & (9)  & (10) \\
\hline
  35 &    12 35 49.44 +62 15 36.9& 2.203 & $<$6.7  & $0.15^{+0.35}_{-0.40}$     & 6.4/8 & $1.75^{+0.44}_{-0.42}$ &  7.8/9 &  42.68 & 1.20 \\
  91 &   12 36 11.40 +62 21 49.9&  0.51 (1.78) & $3.0^{+3.2}_{-2.7}$  & $0.30^{+0.62}_{-0.50}$    &   31.2/17 & $1.74^{+0.26}_{-0.54}$ &  40.3/19 &  42.65 & 9.44 \\
  92 &    12 36 11.80 +62 10 14.5&     0.91 & $4.6^{+8}_{-4.6}$   & $-0.14^{+0.94}_{-0.86}$  & 17.2/7 & $1.16^{+0.44}_{-0.66}$ &  23.2/7 &  42.42  & 2.69 \\
  107 &   12 36 15.83 +62 15 15.5&    0.59 (2.48) & $7.4^{+7.5}_{-6.3}$  & $0.66^{+0.37}_{-0.46}$   & 11.0/12 & $1.93^{+0.21}_{-0.21}$ &  11.1/13 &  43.08  & 1.85 \\
  121 &    12 36 19.89 +62 19 10.1&  0.520 & $0.3^{+2.1}_{-0.2}$  & $0.49^{+0.37}_{-0.43}$   & 4.8/8 & $2.8^{+0.37}_{-0.40}$ &  11.7/9 &  41.93  & 1.90 \\
  127 &    12 36 21.21 +62 11 08.8&  1.014 & $12.3^{+2.5}_{-9.0} $  & $0.20^{+1.20}_{-0.70}$   & 12.9/12/8 & $1.16^{+0.23}_{-0.31}$ &  13.2/13 &  42.76&  3.98  \\
  135 &     12 36 22.66 +62 16 29.8& 2.466 & $275^{+195}_{-138} $  & 1.8   & 5.5/4  &  -  &  - & 43.61  & 1.14 \\
  171 &     12 36 32.59 +62 07 59.8& 1.993 & $ 2.2^{+15}_{-2}$  & $-0.02^{+0.75}_{-0.47}$   & 7.2/9 & $1.65^{+0.31}_{-0.36}$ &  6.2/10 &  42.73  & 1.83 \\
  190 &     12 36 35.58 +62 14 24.1& 2.005 & $1.5^{+1.0}_{-1.5} $  & $0.28^{+0.51}_{-0.29}$  & 13.2/13 & $1.93^{+0.20}_{-0.20}$ &  11.2/14 &  43.01  & 2.48 \\
  191 &    12 36 35.86 +62 07 07.7& 0.28 (0.93) & $ <20$  & $-1.4^{+1.40}_{-0.47}$   & 38.1/18 & $0.20^{+0.44}_{-0.78}$ &  27.4/16 &  41.73  & 5.35 \\    
 
\hline

\end{tabular}
\begin{list}{}{}
\item The columns  are: (1) Alexander ID number; (2) X-ray Equatorial Coordinates (J2000) (3)  redshift ;  three and two decimal numbers 
 refer to spectroscopic and photometric redshift respectively; in brackets we quote the possible X-ray spectroscopic redshift;  (4) Intrinsic column density in  units 
 of $10^{22}$ \cunits for the power-law model;  (5) photon index in the case of the power-law model; 
(6)  $\chi^2$ and degrees of freedom for the power-law model; 
 (7) Photon index in the case of  the reflection model; 
(8)  $\chi^2$ and degrees of freedom for the reflection model
; (9) Logarithm of the observed luminosity in the 2-10 keV band in units of \lunits 
 (10) observed flux in the 2-10 keV band in units of $10^{-15}$ \funits 
\end{list}
\end{center}
\end{table*}

\subsection{Compton-thick selection criteria}

Here, we describe the X-ray spectral  method used to select the Compton-thick sources. 
 The most secure way to identify a source as Compton-thick is through the detection of the 
  absorption turnover. For a Compton-thick source with a column density   $\sim 10^{24}$ \cunits 
   the absorption turnover occurs at rest-frame energies somewhat higher than 6 keV while at 
    a column density of $5\times 10^{24}$ \cunits   the turnover   occurs at 
     an energy of about 20 keV (e.g. Yaqoob 1997). 
    This implies that owing to the limited effective area of Chandra at high energies,  
     it is very difficult  to detect the absorption turnover even for 
       marginally Compton-thick sources at low redshift. However, at higher redshifts the 
      turnover shifts progressively at low energies, because of the K-correction effect, 
       making the identification of Compton-thick AGN more straightforward. 
        For example,  for a Compton-thick source with a column density of  
         $\rm N_H=10^{24}$ \cunits, the absorption turnover would shift 
          to energies about 2 keV at a redshift of z=2, an energy region where Chandra has large effective area. 
           
           Even at this 
           high redshift the detection of the absorption turnover for a source 
            with five times more column density would be challenging.   
      Fortunately, the Compton-thick sources  give another hint for their presence. 
       Their spectrum is reflection-dominated presenting a very flat  photon index 
        with $\Gamma < 1 $ (eg George \& Fabian 1991, Matt et al. 2000). 
      Then the goal is to discriminate between intrinsically flat sources 
       and those which appear flat because of strong ($> 10^{23}$ \cunits) 
        absorption. Obviously, the use 
        of hardness ratios is completely pointless in this task.  
     Only the use of high quality X-ray spectra can ensure that we can determine with 
 certainty whether a source is a Compton-thick candidate. 
 
 Our methodology is  summarised in the following criteria:
 
 1) The detection of an absorption turnover  translating to a column density of $N_H>10^{24}$ \cunits or alternatively 
 
 2) The detection of a flat spectral index $\Gamma < 1.4$ at a statistically significant level (90\% confidence)
  i.e. the 90\% upper limit of the photon index should not exceed $\Gamma=1.4$.    
  This is an arbitrary selected, albeit extremely conservative, cut-off. We note that the average spectrum of Seyfert galaxies 
   is $\Gamma=1.95$ with a dispersion of  only $\sigma=0.15\pm0.04$ (Nandra  \& Pounds 1994).   
   
   We note that the first criterion can obviously be applied only on the 171 sources (out of 222) with 
    spectroscopic or photometric redshift available.  Therefore, we can be incomplete on our estimates  
     of transmission dominated Compton-thick sources.  
     In contrast, the second criterion (reflection-dominated sources) is 
     applied on all the sources, i.e. regardless on whether there is a redshift available. 

\subsection{The  Compton-thick candidates}

The above method yields 10 candidate Compton-thick sources. The X-ray 
 spectral fit details  are given in table \ref{tab_ct}. 
 One source (135) at a (spectroscopic) redshift of z=2.466 is a transmission dominated 
  Compton-thick source i.e. has been characterised as Compton-thick  
  on the basis of an  absorption turn-over  in its X-ray spectrum.   
 Therefore this can be rather considered as a bona-fide Compton-thick AGN.
  
\begin{figure*}
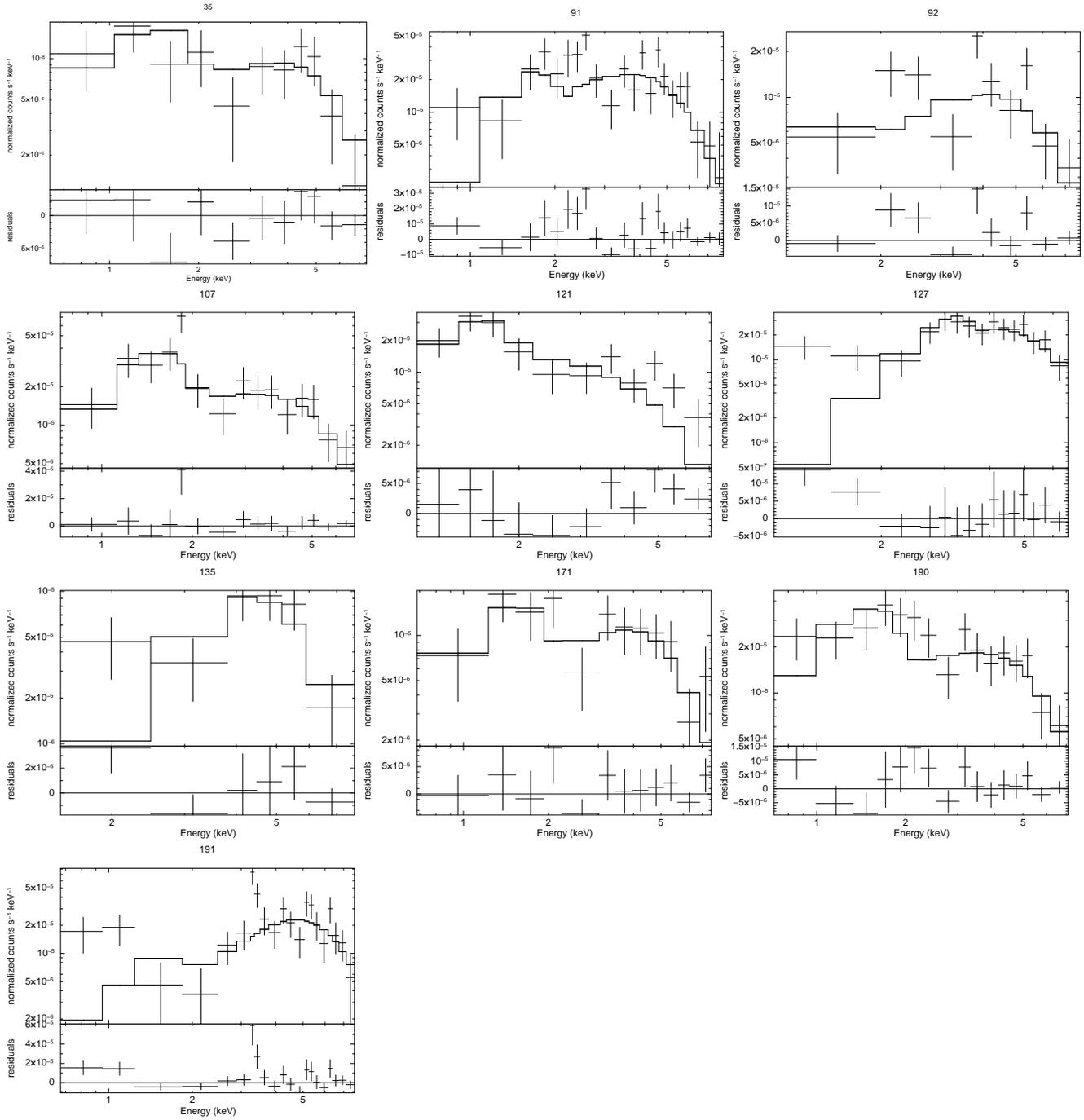

\rotatebox{270}{\includegraphics[width=4.55cm]{35.ps}}
\rotatebox{270}{\includegraphics[width=4.5cm]{91.ps}}
\rotatebox{270}{\includegraphics[width=4.5cm]{92.ps}}\hfill \\
\rotatebox{270}{\includegraphics[width=4.5cm]{107.ps}}
\rotatebox{270}{\includegraphics[width=4.5cm]{121.ps}} 
\rotatebox{270}{\includegraphics[width=4.5cm]{127.ps}}\hfill \\ 
\rotatebox{270}{\includegraphics[width=4.5cm]{135.ps}}
\rotatebox{270}{\includegraphics[width=4.5cm]{171.ps}}
\rotatebox{270}{\includegraphics[width=4.5cm]{190.ps}}\hfill \\ 
\rotatebox{270}{\includegraphics[width=4.5cm]{191.ps}}
\caption{Power-law fits  
 to the spectra of the 10 candidate Compton-thick AGN.}
 \label{xspectra}
 \end{figure*}

  The remaining nine sources present a very flat spectral index suggestive of a reflection continuum. 
   The X-ray spectra together with the single power-law models are shown in Fig. \ref{xspectra}. 
   For the nine flat spectrum sources, 
   we also fit a reflection model, Magdziarz \& Zdziarski (1995), ({\sl pexrav} in XSPEC notation). 
  The best-fit spectral parameter $\Gamma$ for the slope of the incident power-law spectrum is given 
   again in table \ref{tab_ct}. From the $\chi^2$ values it is evident that the reflection model 
    provides an equally good fit to the spectra of most sources. 
     In three cases (sources 91, 92, and 191) 
 neither the single power-law model nor the reflection model can provide a good fit to the data. 
  
  As strong Fe lines are often observed in Compton-thick AGN in the local Universe,
   we further examine whether the addition of a Gaussian component 
   is  required by the data. We choose to adopt here the Cash statistic on the unbinned  spectra.
    In this manner we can obtain more sensitive limits on the EW of the FeK$\alpha$ emission lines. 
       In the case of the source 190,  the addition of a Gaussian component 
   at 6.4 keV rest-frame energy is required by the data   (see table \ref{lines}).  
   In three other cases (91, 107 and 191) a line is required but not 
    at the 6.4 keV rest-frame energy. As all these three sources have only 
     photometric redshifts available it is possible that these features 
      indeed correspond to the FeK$\alpha$ lines. The redshifts inferred  
       by the X-ray spectra for the sources 91, 107 and 191 would then be 1.78,  2.48 and 0.93 
        respectively. We note that the {\it HST} imaging  suggests that source 191 is confused in the Subaru optical images 
         and therefore  its photometric redshift is most probably erroneous. 
  
  One issue which needs to be addressed relates to what is the probability for a reflection dominated flat spectrum 
   to be confused with a highly absorbed spectrum. 
   We have performed spectral simulations with {\sl XSPEC} in order to check this possibility. 
    Full details are given in the Appendix.   
 Our results show that at the faintest fluxes probed here this could be the case. 
  In particular we find that less than 0.2 sources out of the 10 Compton-thick candidates could be spurious.  
 Moreover, we have checked the possibility where a reflection-dominated source with a flat spectrum 
  could evade our $\Gamma <1.4$ criterion. We find that this is highly unlikely with a  probability which is
   lower than 0.1\%.    

%\begin{table*}
%\begin{center}
%\caption{X-ray Lines}
%\label{lines}
%\begin{tabular}{lccccc}
%\hline
%ID & EW & $\chi^2$ &  E & EW & $\chi^2$ \\
 %   (1) & (2) & (3) & (4) & (5)   \\
%\hline 
%35 & $<0.98$  & 6.1/7 & -  & - &  - \\
%91 & $<0.72$ & 30.7/18  & 2.3 & $0.51^{+0.25}_{-0.40}$ & 25.5/17 \\
%92 & $<0.80$ & 16.9/16 & - & - & - \\ 
%107 & $<0.30$  & 11.0/11 & 1.84  &  $0.37^{+0.37}_{-0.27}$ &  4.9/10 \\
%121 & $<0.60$  & 4.8/7 & -  & - &  - \\
%127 & $<0.75$  & 10.8/11 & -  & - &  - \\
%135 & $<6$  & 4.0/2 & -  & - &  - \\
%170 & $<1$  & 6.9/8 & -  & - &  - \\
%190 & $0.51^{+1}_{-0.43}$  & 9.7/12 & -  & - &  - \\
%191 & $<0.20$  & 38.2/17 & 3.32 & $0.91^{+0.43}_{-0.28}$ &  24.8/16 \\
%\hline 
%\end{tabular}
%\begin{list}{}{}
%\item The columns  are: (1) Alexander ID number; (2) EW in keV of Fe line at rest-frame 6.4 keV (3) $\chi^2$ and degrees of freedom   (4) Line Enegy at observer's frame (5) EW in keV (6) $\chi^2$ and degrees of freedom   
%\end{list}
%\end{center}
%\end{table*}

\begin{table}
\begin{center}
\caption{X-ray Lines}
\label{lines}
\begin{tabular}{lcc}
\hline
ID & z & EW   \\
  (1) & (2) & (3)    \\
\hline
35   & 2.203   &  $<490$             \\
91   & 1.78$^{\star}$   &  $166^{+251}_{-136}$ \\
92   & 0.91    &  $<369$             \\
107  & 2.48$^{\star}$   &  $278^{+317}_{-195}$ \\
121  & 0.520    &  $<291$             \\
127  & 1.014  &  $<550$            \\
135  & 2.466    &  $<2800$            \\
171  & 1.993    &  $<732$           \\
190  & 2.005    &  $237^{+479}_{-210}$  \\
191& 0.93$^{\star}$    &  $590^{+422}_{-273}$  \\
\hline 
\end{tabular}
\begin{list}{}{}
\item The columns  are: (1) Alexander ID number; (2) redshift; an asterisk denotes X-ray redshift (3) EW in eV of Fe line at rest-frame 6.4 keV 
\end{list}
\end{center}
\end{table}

%%%%%%%%%    X_RAY SPECTRA %%%%%%%%%%%%%%%%%%%%%%%%

 %%%%%%%%%%%%%%%%%%%%%%%%%%%%%%%%%%%%%%%%%%%%%

\subsection{IR properties} 
The mid-IR data can provide additional diagnostics for Compton-thick AGN.
This is because the heated dust from the AGN radiates copiously at mid-IR wavelenghs 
 (eg Lutz et al. 2004, Alexander et al. 2005). 
  The SED fitting can then immediately give a tentative AGN classification 
   depending on whether the presence of hot dust ($>$300 K)  is required.  The 
    IR fluxes and classification are summarised in table \ref{ir}. 
     Only sources 171 and 190 are classified as AGN while the other sources 
      with available mid-IR data are best-fit with star-forming galaxy templates.

%%%%%%%%%%%%%%%%%%%%%%%%%%%%%%%%%%%%%%%

   \begin{figure*}
   \begin{center}
\includegraphics[width=11.5cm]{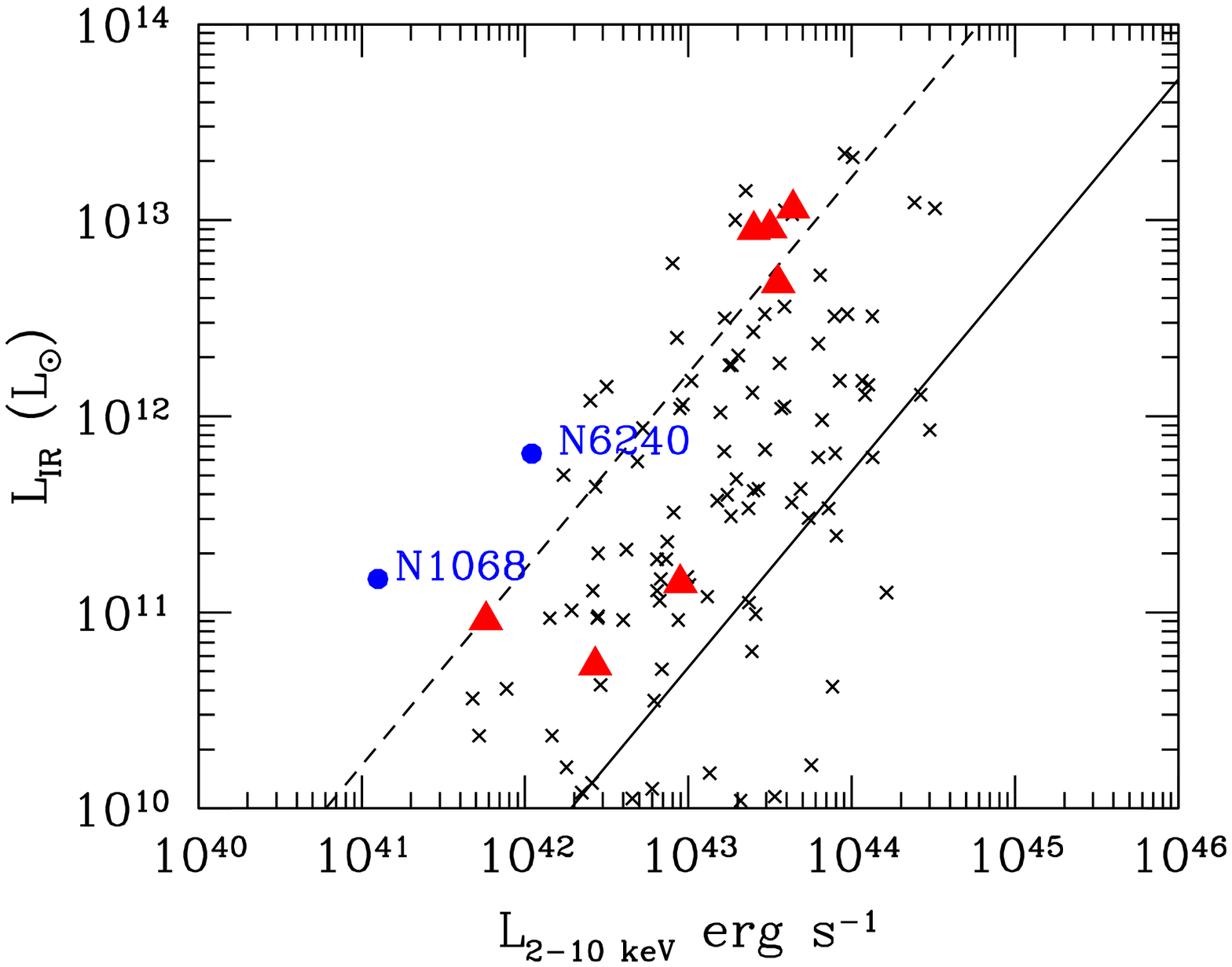} 
\caption{ The total IR luminosity against the observed X-ray (2-10 keV) luminosity for the 
Compton-thick sources (red triangles) and the other X-ray sources (crosses). The solid line 
 corresponds to the average IR to X-ray luminosity expected for an unabsorbed 
 AGN while the dashed line corresponds roughly that expected in the case of a  
  Compton-thick source (see text). The blue circles denote the nearby 
   Compton-thick AGN, NGC1068 and NGC6240.}
  \label{lxlir}
  \end{center}
\end{figure*}

%%%%%%%%%%%%%%%%%%%%%%%%%%%%%%%%%%%%%%%%%    

  The ratio of the observed X-ray to IR luminosity can provide further 
   clues on whether a source is obscured. The ratio of the X-ray 
    to the IR luminosity should be suppressed in the most obscured sources 
     (eg Alexander et al. 2005).     
  We present the  total IR     luminosity
   against the 2-10 keV absorbed luminosity in Fig. \ref{lxlir}.   
    The total IR luminosity is derived using our SED fitting
    model. Note that four of our candidate Compton-thick sources are outside 
 the area covered by the {\it Spitzer} survey. In one case (35), we quote instead the 
  IR luminosity given by Chapman et al. (2005). The solid line denotes the area
   populated by unabsorbed AGN, while the dashed line   indicates the area populated by 
     Compton-thick sources. In the case of Compton-thick sources we assumed that the  reflected emission
      represents 3\% of the intrinsic emission (e.g. Comastri 2004, Akylas \& Georgantopoulos 2009).      
  Most sources show a low ratio of X-ray to IR luminosity consistent with being  Compton-thick.
   We note that although normal galaxies populate the same space of the X-ray/IR diagram,  
   none of our sources can be associated with such objects. This is because
    normal galaxies have steep spectra and low X-ray luminosities 
     (e.g. Georgakakis et al. 2006, Tzanavaris \& Georgantopoulos 2008 and references therein).  
   At least one source (107) appears to have a high X-ray to IR flux ratio 
      casting doubt on its classification as a Compton-thick source.  
       Intriguingly, this source is  an 24$\mu m$ luminous, optically faint source 
        and thus in principle should be an excellent candidate for being obscured (see Fig. \ref{dey}).    
   Four of our candidate Compton-thick sources  have high total IR luminosities 
   ($\rm >10^{12} L_\odot$) lying in the Ultraluminous IRAS Galaxy (ULIRG) regime
    (Sanders \& Mirabel 1996).  All these are sub-mm luminous
     galaxies (Chapman et al. 2005, Alexander et al. 2005) at a redshift higher than two.

%%%%%%%%%%%%% LIR TABLE %%%%%%%%%%%%%%%%%%            
\begin{table*}
\begin{center}
\caption{IR properties of the X-ray selected Compton-thick AGN}
\label{ir}
\begin{tabular}{lccccccccc}
\hline 
ID  & z &  $3.6\mu m$      & $4.5 \mu m$   & $5.8\mu m$ & $8.0 \mu m$   & $24\mu m$  & $R-[24]$  & $log L_{IR} $ & Comments  \\ 
(1)   & (2)&    (3)        &  (4)          &  (5)         &   (6)       & (7)        & (8) & (9) &  (10)  \\
\hline
  35 & 2.203 &  - & - & - & - & - & - & $12.95^{\dagger} $ & sub-mm, oSFR     \\
  91 & 0.51 &  -   &   - & -  & -  & -  & -   &      -     &   \\  
  92 & 0.91 &  17.82 &  17.07 & 16.64 &  16.27 &  160.6 &  14.16  &  11.15 & irAGN\\ 
  107 & 0.59 & 18.54& 17.49&  16.47 & 15.45& 325.3 & 16.1  &  10.73 & irSFR\\ 
  121 & 0.520 &  -   &   - & -  & -  & -  & -   &      -     &   -   \\  
  127 & 1.014 &   -   &   - & -  & -  & -  & -   &         -  &  \\  
  135 & 2.466 &  17.91 & 17.05 & 16.31& 15.97 & 376.4 & 13.88 & 11.6 & sub-mm, oSFR, irSFR \\
  171 & 1.993 &  17.86 & 17.09 & 15.92 & 14.38 &  801.1 7  & 13.60  & 12.96 & sub-mm, oAGN, irAGN\\ 
  190 & 2.005 & 16.60 & 15.64 & 14.53 & 13.33 & 1426.0 & 14.32 &  13.06 & sub-mm, oAGN, irAGN \\   
  191 & 0.276 & 16.08 & 15.81 & 15.55 &  14.77 &  -    &   -    & 10.96 & irSFR\\
 
\hline

\end{tabular}
\begin{list}{}{}
\item The columns  are: (1) Alexander ID number; (2):  redshift ;  three and two decimal numbers 
 refer to spectroscopic and photometric redshift respectively;  (3) 3.6$\mu m$ flux in  units 
 of $\mu Jy$ ;  (4) 4.5$\mu m$ flux in  units  of $\mu Jy$ ;    
 (5) 5.8$\mu m$ flux in  units of $\mu Jy$ ; 
 (6) 8.0$\mu m$ flux in  units of $\mu Jy$ ;  
(7)  24$\mu m$ flux in  units of $\mu Jy$ ; 
 (8) R-[24] colour (Vega); (9) Logarithm of the total IR luminosity in units of $L\odot$  
 (10) Comments: sub-mm denotes that the source is a sub-mm galaxy in the sample 
  of Chapman et al. 2005;  irAGN or  irSFR denotes whether the source has been classified as a 
   star-forming galaxy or an AGN according to the IR-SED fit; oAGN or oSFR denotes that the source 
    has been classified as an AGN or a starburst according to the optical spectrum of Chapman et al. 2005 
  $\dagger$: Luminosity from Chapman et al. 2005  
\end{list}
\end{center}
\end{table*}

\section{Discussion} 

\subsection{X-ray properties}
Our X-ray spectroscopy method revealed 10 candidate Compton-thick sources. 
Only one is a transmission object while the other 9 are reflection-dominated. 
The transmission object at a spectroscopic redshift of z=2.466  can be considered as the more 
 robust Compton-thick case as we see  the spectral curvature directly.
 Its column density is $\sim3\times 10^{24}$ \cunits. 
 The most distant reflection-dominated source in our sample is at a redshift of z=2.2. 
  At this redshift the column density should exceed $\sim 5\times10^{24}$ \cunits
   so that the transmitted emission is not redshifted and thus observable in the \chandra passband. 
   This would be then comparable  to the the highest column densities 
    observed in the local Universe (e.g. NGC4945, Itoh et al. 2008).  
  
  \subsubsection{Fe lines}
   The presence of  high equivalent width $\rm FeK\alpha$ lines  should provide 
    conclusive evidence on whether our sources are Compton-thick. 
  High equivalent width  $\rm FeK\alpha$ lines  are present in the two  
  nearest Compton-thick AGN  (NGC1068, Matt et al. 2004; Circinus, Yang et al. 2009).
   This line is often considered to be the 'smoking gun' revealing the presence 
    of a heavily obscured nucleus.  
   In our sample,  there is no clearcut evidence for the presence of a 
    {\it strong} ($>$1 keV)  $\rm FeK\alpha$ line which would unambiguously classify 
    a source as a bona-fide Compton-thick AGN.  
   We clearly detect a $\rm FeK\alpha$ line  in the case of the source 190
     (but with an EW 90\% upper limit of  only $\sim$700 eV).   
   In three cases we are observing features which are consistent with strong emission lines
    but not at the 6.4 keV rest-frame energy. As all three sources have only photometric redshift available,  
     it is possible that these features correspond indeed to $\rm FeK\alpha$ lines. 
   The EW in all the above cases present large uncertainties, but interestingly 
    none exceeds 1 keV at the 90\% upper limit.  
     In other cases, the derived upper limits are consistent with high 
      EW (e.g. sources 135, 170) while in the case of the sources 35, 92 and 121, 
       the 90\% upper limits are low ($<$500 eV).
       
\begin{figure*}
\rotatebox{270}{\includegraphics[width=8.5cm]{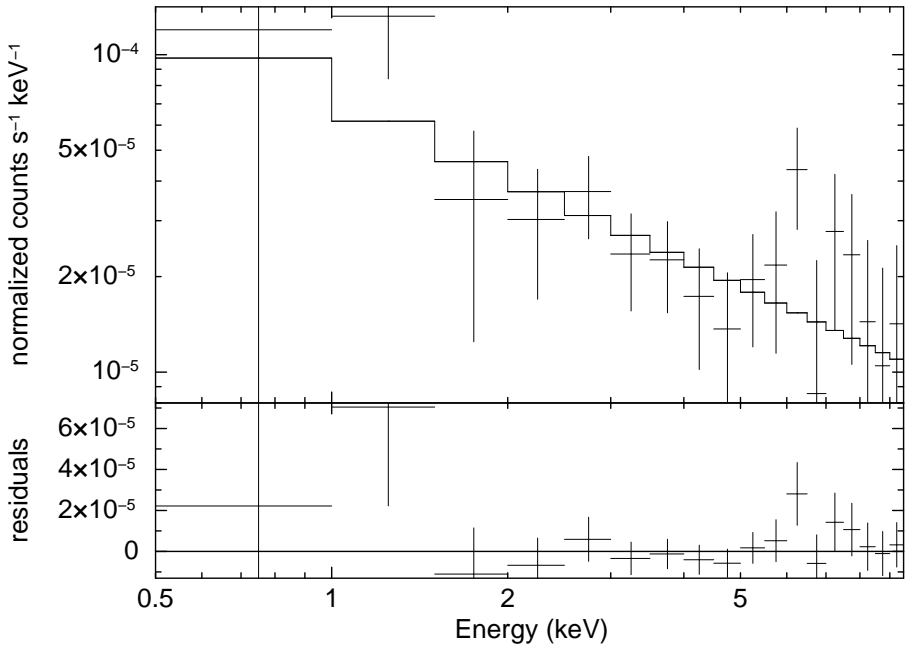}}
\caption{The stacked X-ray spectrum of the six candidate Compton-thick sources
 with spectroscopic redshift available.}
\label{stack}
\end{figure*}

       We have attempted to coadd the spectra in order to put more stringent 
        constraints on the presence of an Fe line. We use only the six objects 
         with  spectroscopic redshift available (see table 1). 
        The addition of the ungrouped spectra has been done in the following way. 
        We first obtain for each source the unfolded spectrum, corrected 
          for instrumental effects,  using {\sl XSPEC}. 
          We then shift each bin to the rest-frame 
           energy according to the redshift of each source.
            At this stage the spectra are stacked. 
             The spectrum is shown in Fig. \ref{stack}.
             The data are grouped in 0.5 keV bins.   
         An FeK$\alpha$ line at 6.4 keV rest-frame is clearly detected, 
           with an EW of $790^{+500}_{-650}$ eV. 
        
       Next, we discuss whether the absence of strong lines argues against 
         the presence of Compton-thick AGN. Murphy \& Yaqoob (2009) 
      present Monte-carlo simulations for Compton-thick reprocessors.
        For a column density of $10^{24}$ \cunits and for a toroidal geometry, they predict EW 
       varying between 100 and 400 eV depending on the viewing angle
        (but see also the models of Ikeda et al. 2009). The EW rise up to 1 keV 
         for column densities $\rm N_H\sim 10^{25}$ \cunits. Therefore,  
          the models of Murphy \& Yaqoob (2009) would tend to suggest that 
           at least some of our objects are not transmission dominated Compton-thick sources.
            Nevertheless, it is possible that there is large variety of obscuring screen geometries 
             resulting in a wide range of FeK$\alpha$ line EW. 
              For example, in the SWIFT/BAT sample of   
     Winter et al. (2009) there is a large number of sources with 
      column densities very close to $10^{24}$ \cunits.
       These have EW ranging within an order of magnitude from about 100 to 1000 eV. 
        This clearly demonstrates the uncertainty in determining the 
         column density from the EW of the FeK$\alpha$ line
          (see also Brightman \& Nandra 2008  for  
           the case of IRASF 01475-0740).

 \subsection{IR, sub-mm and optical properties}
  Four of our Compton-thick AGN are bright sub-mm sources providing further indirect 
  evidence that these may host heavily buried nuclei. These include the transmission 
   dominated source. 
   Alexander et al. (2005) independently argued that these sources 
    host Compton-thick nuclei. These authors discuss 
    the properties of the common X-ray/sub-mm sources in the CDFN. They find that 
   11 X-ray sources coincide with sub-mm sources down to the X-ray flux level
    adopted here i.e. $\rm f_{2-10~ keV} = 10^{-15}$ \funits.   
    Good quality optical spectra are available 
   for the four sub-mm sources from Chapman et al. (2005). 
    As discussed by these authors, only two of these show AGN signatures 
    while the other two are consistent with star-forming galaxy spectra.  
     Pope et al. (2008b) and Menendez-Delmestre (2009) present {\it Spitzer} IRS spectra of 
      sub-mm galaxies. These include our Compton-thick candidate sources 135
      and 35. The IRS spectra of these two sources appear to be dominated by  star-forming emission 
   i.e. flat continua, PAH emission features. They also present strong silicate absorption features
    suggestive of large column densities. 
          
\begin{figure}
\includegraphics[width=8.5cm]{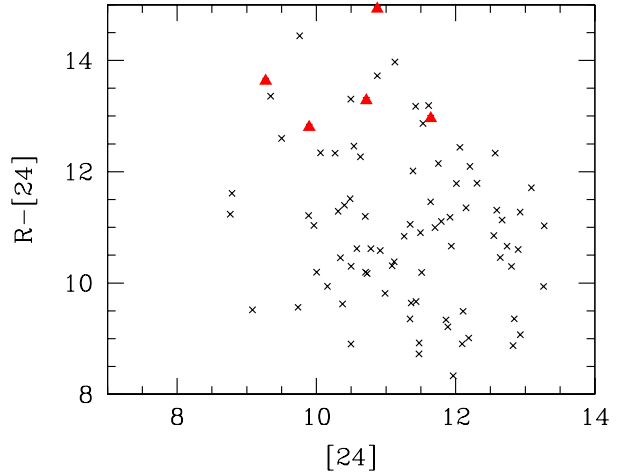} 
\caption{ The R-[24] colour vs. [24] magnitude for the 5 candidate Compton-thick sources (red triangles),
compared with the other X-ray sources with  24$\mu m$ Spitzer MIPS detections (black crosses).}
\label{dey}
\end{figure}

    The {\it Spitzer} mission revealed a class of bright Infrared, optically faint  sources 
    (Houck et al. 2005). These have extreme 24$\mu m $ to R-band flux ratios 
     $\rm f_{24}/f_R\sim 1000$ or $R-[24] > 14$  (Dey et al. 2008). 
     These are often referred in the literature as Dust  Obscured galaxies (DOGs). 
     Fiore et al. (2008) propose that the majority of these sources are associated with 
      Compton-thick sources at high redshift. Our five Compton-thick candidate sources 
        with available {\it Spitzer}   mid-IR data,  have very high mid-IR to optical flux values:
         $R-[24]  > 13.6$.  In Fig. \ref{dey} we give the R-[24] magnitude for our candidate Compton-thick sources 
         in comparison with the rest of our sample. The candidate Compton-thick sources 
          populate the upper (i.e. redder or optically faint) part of the diagram. 
           This provides additional support to the claims which link the infrared bright, optically faint sources
            (DOGs)  with Compton-thick AGN. 
             Pope et al. (2008) assert that the DOG population is  substantially contaminated
              by star-forming galaxies.   
         These authors have proposed a diagnostic diagram for DOGs. 
          In particular they find that the DOGs which are associated with 
           star-forming galaxies, according to {\it Spitzer} IRS spectroscopy, have a flux ratio 
            $S_{8.0}/S_{4.5}<2$, while the AGN present higher values.  This is not the case in our sample. 
             From table 3, we see that all six candidate Compton-thick AGN  with available {\it Spitzer} data
              have $S_{8.0}/S_{4.5}<1$ and thus would be  classified as star-forming galaxies 
               according to the above criterion.    
 
\begin{figure}
\includegraphics[width=8.5cm]{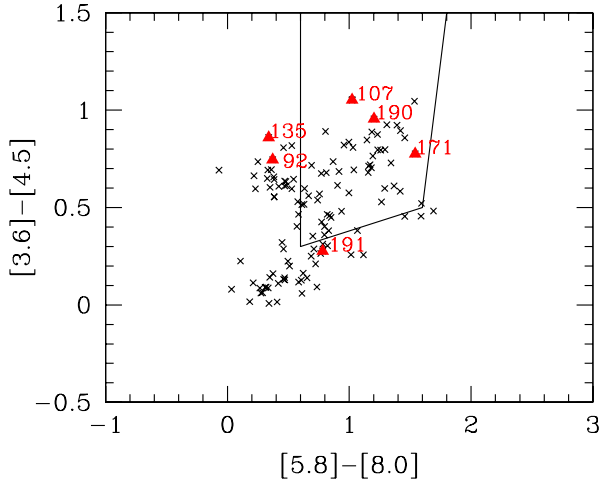} 
\caption{The mid-IR colour-colour diagram for the Compton-thick sources 
(red triangles) as compared with the other X-ray sources in our sample (black crosses). The 'wedge'
 defines the region occupied by mid-IR selected AGN (Stern et al. 2005)}
\label{stern}
\end{figure}

 It is instructive to examine the location of our candidate Compton-thick sources on the 
     {\it Spitzer} IRAC mid-IR colour-colour diagram. 
   Stern et al. (2005) have demonstrated that this diagram provides an efficient 
    diagnostic for AGN identification. In particular, AGN are found to occupy the 'red' part of the 
   [3.6]-[4.5] vs [5.8]-[8.0]  band   diagram. The power-law IR spectrum 
    sources of Donley et al. (2007) or Alonso-Herrero et al. (2007) would also 
     fall in this 'wedge'. 
    We plot the colour-colour diagram in Fig.\ref{stern}. From the five candidate Compton-thick 
      sources only three (107, 171, 190) appear to be inside the AGN wedge. 
       One source (191) is lying just below the 
      wedge  on a region occupied by normal galaxies (see e.g. Barmby et al. 2006). 
       Finally two more sources (135 and 92) are having [5.8]-[8.0] colours bluer than the AGN.
        This is the part of the diagram which is populated by the mid-IR bright, optically faint sources 
        (see Georgantopoulos et al. 2008).  
      It appears then that the IR AGN selection methods based on either colours or 
       power-law spectra would fail to detect a large fraction of Compton-thick candidates. 

 Finally, we examine the relation of our candidate Compton-thick sources  with the 
  extreme X-ray to optical flux ratio sources (Koekemoer et al. 2004). 
   These sources are believed to be associated with heavily obscured AGN at high redshift. 
    This is because at high redshift the k-correction shifts the optical wavelengths to the UV making them 
     more prone to dust absorption while in contrast the k-correction shifts the X-ray wavelengths 
      to higher energies which are less obscured. This results in an increase of the X-ray to 
       optical flux ratio for absorbed sources with increasing redshift.   
  Civano et al. (2005) present the X-ray spectral properties of the optically faint hard X-ray sources in 
   the CDFN. 
    They find 63 sources with extreme X-ray to optical flux ratios ($\rm \log (f_x/f_o)>1$). 
    According to Civano et al. (2005), the stacked X-ray spectra of these sources suggest that these are heavily obscured sources. 
    Only three sources coincide with the Compton-thick candidates here (sources 91, 92 and 107) suggesting 
     that the extreme X-ray to optical flux ratio method may not be highly efficient for selecting out 
      Compton-thick sources.

\subsection{Comparison with X-ray background synthesis models}  
\begin{figure}
\includegraphics[width=8.5cm]{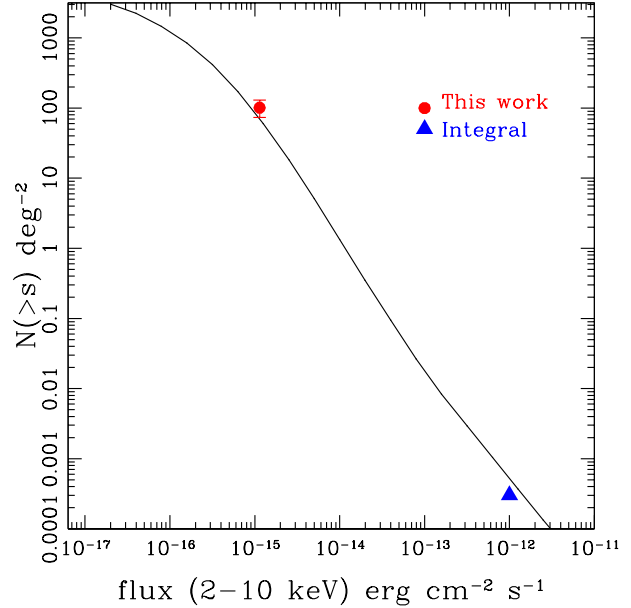} 
\caption{ The number counts for the 10 candidate Compton-thick sources (red circle) 
compared  with the predictions (solid line) of the X-ray background synthesis model of
 Gilli et al. (2007). The blue triangle denotes the {\it INTEGRAL}  number counts (adapted from Treister et al. 2009).
 The dash line corresponds to the total logN-logS.}
\label{lognlogs}
\end{figure}

 Here, we compare the flux distribution as well as the 
  redshift distribution of the 10 candidate Compton-thick AGN 
    with the predictions of the X-ray background 
 synthesis models of Gilli et al. (2007). We use the POMPA code
 \footnote{www.astro.bo.it/\~{}gilli/counts.html}. We assume that Compton-thick AGN have column 
  densities up to $10^{26}$ \cunits.    The predicted number count distribution logN-logS 
   is given   in Fig. \ref{lognlogs}.  
    If we assumed only sources with column densities 
   $<10^{25}$ \cunits, we obtain roughly a factor of two lower normalization in the number counts.
   Our point agrees well with the logN-logS of Gilli et al. (2007).
    We also plot the observed number density  at bright fluxes derived from the 10 Compton-thick sources of {\it INTEGRAL}  (Sazonov et al. 2007),
     as adapted from Treister et al. (2009). The conversion from the 17-60 keV to the 2-10 keV band has been performed 
 assuming a   spectrum of $\Gamma \approx 0.2$, i.e. the median spectrum of our sources. 
    Treister et al. (2009) point out that the {\it INTEGRAL} number counts are more than a factor of two below the predictions of the Gilli et al. model. This could be partially explained by the fact  
    that the {\it INTEGRAL} surveys cannot detect sources with $\rm N_H>10^{25}$ \cunits, assuming of course that
     such heavily obscured sources really exist. However, Treister et al. assert that 
      one can obtain a very good fit to the X-ray
     background spectrum  in the 20-40 keV energy range without the need for
     a large number of  heavily obscured sources.   This is because of the uncertainties on the 
     normalization of the X-ray background at high energies as well as on the uncertainty on     
    the AGN spectrum at high energy i.e. the normalization of the reflection component.      
  
    The number count distribution of the Compton-thick population illustrates very well 
     why the Chandra Deep Fields offer the most promising area for detecting 
     Compton-thick AGN, at high redshift, through X-ray spectroscopy. At the flux limit imposed here 
      ($10^{-15}$ \funits), we barely accumulate a few hundred photons in order 
       to have a sufficient number of  counts for spectral fitting. For a survey with 
        10 times less exposure time (similar to the COSMOS survey Hasinger et al. 2007;  or 
         the AEGIS survey Nandra et al. 2007), 
         we would obtain the same number of photons for sources with 
          a flux of  about $10^{-14}$ \funits. However, the number density at these bright fluxes 
      is only  $\rm \sim 1 deg^{-2}$ i.e.  about two orders of magnitude lower.           
      
      \begin{figure}
\includegraphics[width=8.5cm]{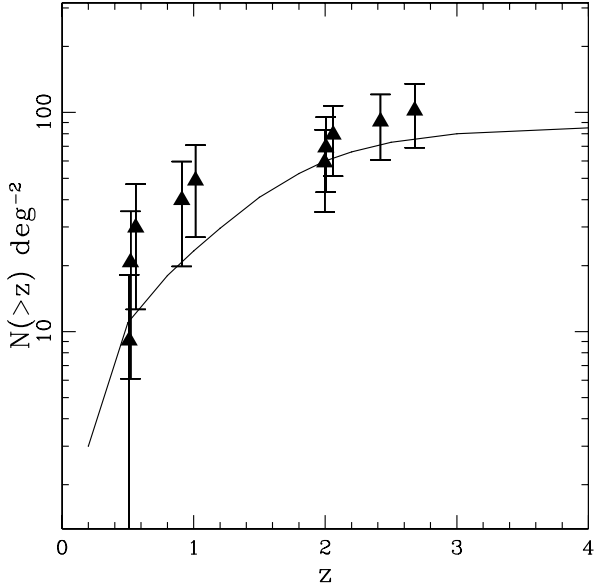} 
\caption{ The number counts for the 10 candidate Compton-thick sources (triangles) 
compared  with the predictions (solid line) of the X-ray background synthesis model of
 Gilli et al. (2007)}
\label{nz}
\end{figure}

     In Fig. \ref{nz} we plot the cumulative redshift distribution for our sample. This is compared with the
     predictions of the Gilli et al. model. There is rather  reasonable overall agreement, given the 
      uncertainties introduced by  the photometric redshifts.   The number of Compton-thick AGN below a redshift 
       of z=0.5 is about 10\% of the total number up to a redshift of z=3. Still these low redshift AGN 
        are the ones which contribute 50\% of the X-ray background. Instead the contribution of the 
         high redshift Compton-thick sources is very small: less than 1\% for those with $z> 2$ 
         (see Treister et al. 2009, Gilli et al. 2007).

 \section{Conclusions} 

We have investigated the X-ray spectral properties of the bright CDF-N sources 
$>10^{-15}$ \funits selected in the 2-10 keV band. There are 
 171 sources which have either 
  a spectroscopic or a photometric redshift available. Our aim is to 
  identify Compton-thick candidates either though the presence 
    of a absorption turnover or a reflection-dominated spectrum. 
    Our conclusions can be summarised as follows:   
    
    \begin{itemize} 
    \item{We find 10 candidate Compton-thick AGN. For six of them 
     there are spectroscopic redshifts while for the remaining 
      only photometric available. One source is a
     transmission Compton-thick AGN while the 
      other nine are probably reflection dominated.}
      \item{We find no high EW ($>1$ keV)
        FeK$\alpha$ emission line in the individual spectra of our sources.
         However, the stacked spectrum of the six sources
          with spectroscopic redshift reveals an FeK$\alpha$ line 
           with an EW of $\approx$800 eV.} 
    \item{Four sources, including the transmission 
     Compton-thick object,  are associated with luminous sub-mm
      galaxies, at redshift $z \sim2$  or higher, further supporting the heavily 
       obscured scenario for these.}    
    \item{For many of the sources with available IR luminosity, we find a low X-ray to 
     total IR luminosity ratio corroborating
     the presence of obscured accretion.} 
    \item{Our Compton-thick candidates appear  to be associated  with optically 
     faint, 24$\mu m$ bright Infrared galaxies in agreement with the findings of Fiore et al. (2008) }
     \item{Interestingly, mid-IR selection criteria (e.g. colour-colour diagram) do not appear to be  efficient in selecting 
      our candidate Compton-thick AGN.  The same applies to methods which use the X-ray 
       to IR or optical flux ratio. Instead, the selection of mid-IR bright and faint optically sources has
        the most significant overlap with our sample}  
    \item{The number counts of Compton-thick sources at faint fluxes are in overall  agreement 
     with the predictions of the population synthesis models of Gilli et al (2007)} 
    \end{itemize}
    
         Future observations with the {\it Nustar}, {\it Simbol-X} and {\it Next} missions will provide 
     the first images of the  hard X-ray Universe at energies above 10 keV. 
     These are expected to provide a wealth of data on Compton-thick sources 
     at low redshift and place tight constraints   on the contribution of these objects 
      to the X-ray background.  In the meantime, XMM-Newton can shed light 
       on the properties of Compton-thick sources at moderate to high redshift. The 
        already scheduled 
        3Ms XMM-Newton observations of the Chandra Deep Field South can 
         provide spectra of unprecedented quality and thus to  easily 
          identify the Compton-thick candidate sources.  
    
\begin{acknowledgements}
We thank the anonymous referee for his useful suggestions.
 We also thank Roberto Gilli and Ezequiel Treister for their 
  comments.   
 We acknowledge the use of {\it Spitzer} data provided by the 
 {\it Spitzer} Science Center. 
 The Chandra data were taken from the Chandra Data Archive 
 at the Chandra X-ray Center.   
\end{acknowledgements}

\begin{appendix}
\section{Spectral Simulations}
We perform spectral simulations in order to quantify 
the fraction of Compton-thick candidates 
that have been erroneously included in our sample. 
 In particular, because of the limited photon statistics, 
   some faint heavily absorbed sources 
  may be mistaken for flat sources. 
  We use XSPEC v.12.4 to create 1000 fake X-ray spectra 
in the 2-10 keV flux interval of $10^{-14} - 10^{-15}$ \funits  
i.e. the flux range of the actual Compton-thick sources.  
The on-axis response and auxilliary files for $\chandra$ ACIS-I CCD
are used for simplicity. We assume an absorbed power-law model with $\Gamma$ 
fixed to 1.9 and a  column density randomly varying
between $10^{20} - 10^{24}$  \cunits.
We fit the fake spectra using XSPEC using an absorbed power-law model with both 
the $\Gamma$ and $\rm N_H$ parameters free. We then  estimate the 
fraction of sources that present a best fit $\Gamma$ value flatter than 
1.4 at the 90 per cent upper limit.
The spectral fits show that  2 out of 1000 sources
are spurious Compton-thick candidates. We note here that these simulated 
source present high columns ($\rm \sim6\times10^{23}$ cm$^{-2}$). Our sample
contains 141 in the same flux interval. This automatically suggest that 
the number of the spurious Compton-thick candidates  in our sample is $\sim$0.3. 
  
Inversely, it is likely that    
   within the errors, some reflection-dominated flat sources  may
    have a 90\% photon index upper limit $\Gamma> 1.4$
    and thus are missed by our criteria.  
In order to check this possibility we run another simulation. 
 We create 1000 fake spectra using the {\sl PEXRAV} 
model in XSPEC. We assume z=1 and a random 2-10 keV flux between  
$10^{-15} -10^{-14} $ \funits. We again fit the spectra in XSPEC 
using an absorbed power-law  model. We estimate the fraction of the 
sources that show a power-law photon index steeper that 1.4 at the 
90 per cent confidence level. Our simulations show that  
there are no sources with $\Gamma$ less than 1.4 at the 90 per cent 
upper limit.  This immediately suggests  that the number of missed 
 Compton-thick sources   in our sample is $<0.14$. 
\end{appendix}

%%%%%%%%%%%%%%%%%%%%%%%%%%%%%%%%%%%%%%%%%%%%%%%%%%%%%%%%%%%%%%%%%%%%%%%%%%
%_____________________________________________________________
%                              Table longer than a single page  
%  In the preamble, use:              \usepackage{aalongtable}
%-------------------------------------------------------------
%          All long tables have to be placed at the end, after 
%                                        \end{thebibliography}
%\begin{longtable}{ccccccc}
%\caption{\label{stacked} Optical and infrared properties of the stacked sources}\\
%\hline \hline
%num & RA(24$\mu$m) & DEC(24$\mu$m) & f$_{\rm R}$ & f(24$\mu$m) & log(L$_{ir}$) & $z$          \\
%    & deg          & deg           & $\mu$Jy     & $\mu$Jy     & L$_{\odot}$   &              \\
%\hline
%\endfirsthead
%\caption{continued.}\\
%\hline\hline

%\label{stacked}
%%
%\end{longtable}
%Notes:
%\begin{enumerate}
%\item{Spectroscopic redshift}
%\end{enumerate}
%
\end{document}